\documentclass[10pt,twocolumn,letterpaper]{article}

\usepackage{iccv}
\usepackage{times}
\usepackage{epsfig}
\usepackage{graphicx}
\usepackage{amsmath}
\usepackage{amssymb}

\usepackage{caption}
\usepackage{subcaption}
\usepackage{svg}
\usepackage{comment}
\usepackage{hhline}
\usepackage{multirow}
\usepackage{booktabs}

\usepackage[pagebackref=true,breaklinks=true,letterpaper=true,colorlinks,bookmarks=false]{hyperref}

\iccvfinalcopy % *** Uncomment this line for the final submission

\def\iccvPaperID{CVAMD-24} % *** Enter the ICCV Paper ID here

\newcommand\footnoteref[1]{\protected@xdef\@thefnmark{\ref{#1}}\@footnotemark}
\newcommand*\samethanks[1][\value{footnote}]{\footnotemark[#1]}
\begin{document}

%%%%%%%%% TITLE
\title{Segmentation-based Assessment of Tumor-Vessel Involvement for Surgical Resectability Prediction of Pancreatic Ductal Adenocarcinoma}
%\institute{Eindhoven University of Technology, Eindhoven 5612 AZ, The Netherlands \and Catharina Ziekenhuis, Eindhoven EJ 5623, The Netherlands \and Philips Research,  Eindhoven AE 5656, The Netherlands \newline
\author{Christiaan Viviers\thanks{\enspace Equal contribution.}\textsuperscript{\hspace{0.15cm},1}, Mark Ramaekers\samethanks\textsuperscript{\hspace{0.15cm},2}, Amaan Valiuddin\textsuperscript{1}, Terese Hellström\textsuperscript{1},\\ 
Nick Tasios\textsuperscript{3}, John van der Ven\textsuperscript{3}, Igor Jacobs\textsuperscript{3}, Lotte Ewals\textsuperscript{2},  Joost Nederend\textsuperscript{2},\\
Peter de With\textsuperscript{1}, Misha Luyer\textsuperscript{2}, Fons van der Sommen\textsuperscript{1}\\
\small \textsuperscript{1}Eindhoven University of Technology, \textsuperscript{2}Catharina Ziekenhuis, \textsuperscript{3}Philips, Eindhoven, The Netherlands
}


\maketitle

%%%%%%%%% ABSTRACT
\begin{abstract}
Pancreatic ductal adenocarcinoma~(PDAC) is a highly aggressive cancer with limited treatment options. This research proposes a workflow and deep learning-based segmentation models to automatically assess tumor-vessel involvement, a key factor in determining tumor resectability. Correct assessment of resectability is vital to determine treatment options. The proposed workflow involves processing CT scans to segment the tumor and vascular structures, analyzing spatial relationships and the extent of vascular involvement, which follows a similar way of working as expert radiologists in PDAC assessment. Three segmentation architectures (nnU-Net, 3D U-Net, and Probabilistic 3D U-Net) achieve a high accuracy in segmenting veins, arteries, and the tumor. The segmentations enable automated detection of tumor involvement with high accuracy (0.88 sensitivity and 0.86 specificity) and automated computation of the degree of tumor-vessel contact. Additionally, due to significant inter-observer variability in these important structures, we present the uncertainty captured by each of the models to further increase insights into the predicted involvement. This result provides clinicians with a clear indication of tumor-vessel involvement and may be used to facilitate more informed decision-making for surgical interventions. The proposed method offers a valuable tool for improving patient outcomes, personalized treatment strategies and survival rates in pancreatic cancer.
\end{abstract}

%%%%%%%%% BODY TEXT
\section{Introduction}\label{sec:intro}

\begin{figure*}
\begin{center}
 \centerline{\includegraphics[width=1.0\linewidth]{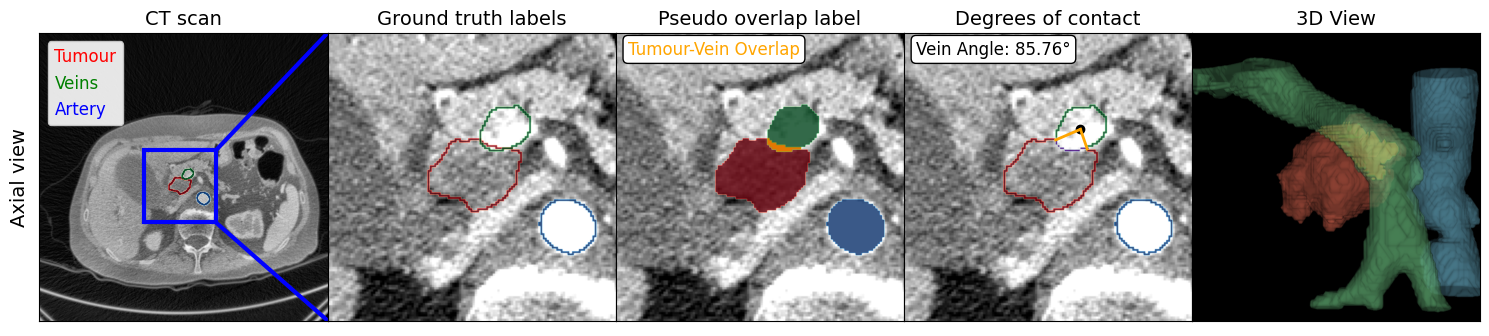}}
\end{center}
\vspace{-1cm}
  \caption{Slice from CT scan depicting the involvement between the tumor and vein, a pseudo overlap label and the computed angle of involvement based on contact pixels (purple). Other labeled structures are omitted.}
  \label{fig:involvement}
\end{figure*}

Pancreatic ductal adenocarcinoma~(PDAC) is one of the most aggressive malignancies with a dismal prognosis and an overall 5-year survival rate of less than 10\%~\cite{Rahib2014-dh}. Despite recent advancements in the field of oncology, pancreatic cancer often goes undetected until it has progressed into an advanced stage. As a result, the majority of patients have advanced or metastatic disease, leading to limited treatment options and poor outcomes~\cite{De_la_Santa2014-jz}. With its high mortality rate and limited treatment options, identifying the optimal management approach for pancreatic cancer patients remains a crucial area of clinical focus. In recent years, the concept of \textit{resectability}~\cite{tummala2011imaging,tempero2021pancreatic,DPCG} has emerged as a pivotal factor in determining the appropriate treatment strategy, emphasizing the importance of accurately assessing the feasibility of curative surgical resection.

Pancreatoduodenectomy~(PD) is the cornerstone for surgical treatment of pancreatic cancer. However, this procedure poses significant technical challenges and is associated with a considerable morbidity rate, ranging from 20\%~\ to 30\%~\cite{Treadwell2016-we}. Furthermore, a mere 20\% of patients are considered eligible for resection upon initial diagnosis~\cite{Hidalgo2010-sp}. Therefore, it is essential to carefully evaluate vascular involvement of the tumor and identify potential arterial anatomical variations during preoperative assessment. These factors play a critical role in determining the feasibility of surgical resection~\cite{de2014radiology}. Currently, multi-phase contrast-enhanced multi-detector computed tomography~(MDCT) is the gold standard for evaluating pancreatic cancer and determining the resectability. Standardized resectability criteria are used to tailor the need for neoadjuvant therapy and select patients for (minimally-invasive) surgical resection~\cite{tummala2011imaging}. Resectability is graded as either \emph{resectable}, \emph{borderline resectable}, or \emph{irresectable}, based on the degrees of contact between the tumor and surrounding vasculature~\cite{tempero2021pancreatic,Preopanc,DPCG}. However, determining surgical resectability based on CT scans can be difficult, especially after neoadjuvant treatment. Tumor regression after neoadjuvant treatment is rarely visible on CT and the amount of vascular involvement tends to be over-estimated~\cite{tempero2021pancreatic,asbun2020miami,alemi2016classification,lermite2013complications}. Moreover, existing literature demonstrates significant interobserver variability, even among highly experienced clinicians~\cite{versteijne2017considerable,joo2019preoperative}. As such, clinicians are hardly able to accurately assess tumor resectability~\cite{tran2014radiographic}.

% To assess surgical resectability, it is critical to determine the \textit{presence} and extent of involvement between the tumor and surrounding blood vessels. This involves acquiring high-quality contrast-enhanced CT scans that provide detailed anatomical information about the tumor and surrounding structures. The tumor and vein/artery segmentations can be performed manually or automatically using advanced algorithms. Analyzing the spatial relationship between the tumor and vascular tree requires identifying points of intersection or contact. The degree of involvement is then computed based on this spatial relationship. Computed angles can be quantified and statistically analyzed to evaluate the extent of tumor-vessel contact, aiding in surgical planning and decision-making.

By leveraging a deep learning-based clinical decision support system~(CDSS), there is a potential for significant improvement in resectability assessment, empowering clinicians with assistance, enhancing the overall accuracy, and reducing interobserver variability of the process. This research proposes a workflow to acquire a focused region of interest concerning PDAC and surrounding anatomical structures. Three deep learning-based segmentation architectures are implemented to segment the structures of interest to ultimately present multiple levels of clinically-relevant information. The initial segmentations are used to asses tumor size and location with respect to the surrounding anatomy. From the tumor and vessel segmentations the involvement is automatically calculated to, first of all, determine \textit{if} there is involvement and secondly, the extent thereof. Each of these steps carries additional clinical value and further insights into patient treatment options. Finally, we present the ambiguity captured by each of the models and show how this ambiguity can aid in the resectability decision-making process. The research contributions are as follows.
\begin{itemize}
   \itemsep0em
  %\item A workflow based on existing deep-learning methods to acquire a focused region of interest in PDAC CT cases enabling resectability assessment.
  \item The implementation and evaluation of 3 deep learning-based segmentation architectures (3D nnU-Net, custom 3D U-Net and Probabilistic 3D U-Net) for the multi-class segmentation of PDAC and relevant surrounding anatomy. 
  \item A new overlap loss~(OLL) function that encourages segmentation of the tumor and vessels in the overlapping regions. 
  \item High segmentation accuracies (veins 0.88 Dice, arteries 0.86 Dice, and importantly, the pancreatic tumor 0.66 Dice), that enable automated detection of tumor-vessel involvement and, where possible, utilizing the segmentation to compute degrees of involvement.
  \item Explainable uncertainty segmentations and visualizations of the critical structures along with direct effects of the uncertainty on the final tumor-vessel involvement assessment.
  \item We are the first to present high sensitivity~(88.2\%) and specificity~(85.7\%) in automated detection and assessment of tumor-vessel involvement, enabling clinicians to make more precise and informed decisions regarding surgical resection.
\end{itemize}
This research paper concentrates on various methods to automatically determine the extent of the tumor-vessel involvement - the core criterion for which the tumor is evaluated for resectability. Automated segmentation of tumor, pancreas, veins, arteries, common bile duct and pancreatic duct is realized using the 3D nnU-Net~\cite{nnUnet}, a 3D U-Net~\cite{3dunet} and the Probabilistic 3D U-Net~\cite{prob3dunet}. The resulting segmentations are employed to compute and predict tumor involvement as a metric for surgical decision-making. The layout and extent of involvement serves as assistance for the surgeon in assessment of resectability and surgical planning.

\section{Related Work on PDAC Detection, Segmentation  \&  Resectability Prediction}\label{sec:related_work}
Deep learning-based methods have demonstrated significant potential in the detection of pancreatic cancer on CT scans. Several studies have employed classification networks and achieved high accuracy in detecting PDAC and other types of pancreatic cancer~\cite{Liu_deep_2020,Vaiyapuri_intelligent_2022,patra_abstract_2021,si_fully_2021,wei_multidomain_2023,wang_maff_2023,liu_establishment_2019,zhang_novel_2020,ma_construction_2020}. Recently, segmentation for the classification of pancreatic cancer has garnered significant attention, since it both detects and localizes cancer~\cite{zhu_multi_scale_2019,wang_learning_2021,zhu_segmentation_2021,alves_fully_2022,chen_pancreatic_2023}. Notably, Viviers~\etal~\cite{viviers_improved_2022} and Alves~~\etal~\cite{alves_fully_2022} have proposed a similar segmentation-for-classification framework, leveraging the surrounding anatomy and secondary tumor indicative features, such as the common bile duct and pancreatic duct, to enhance tumor segmentation and improve detection accuracy. Obtaining an automated detailed segmentation map of the tumor provides high clinical value. As such, Mahmoudi~~\etal~\cite{mahmoudi2022segmentation} have proposed a hybrid 2D-3D segmentation-based approach for detailed segmentation of the tumor mass and surrounding vessels in tumor-only cases. While they showcase good segmentation accuracy (Dice: 0.61 PDAC, 0.81 Artery and 0.73 Vein), they note that a full 3D method will further improve results and will be essential for determining tumor-vessel involvement. Recently, Yao~\etal~\cite{Yao2023-mp} presented a multicenter, retrospective study in which they construct an imaging-derived prognostic biomarker, dubbed DeepCT-PDAC, for overall survival~(OS) rate prediction. They train segmentation~(nnU-Net) and prognostic models (CE-ConvLSTM and Tumor-vascular Involvement 3D CNN) to model the tumor-anatomy spatial relations. The 3D predictions of PDAC, the portal vein and splenic vein (PVSV), superior mesenteric vein (SMV), superior mesenteric artery (SMA) and truncus coeliacus (TC) are used in the Tumor-vascular Involvement 3D CNN branch. Contact area features are predicted to be used in a final risk score or OS prediction. While the research presents impressive results for the accuracy of OS predictions, intermediate steps leading to the final prediction remain unclear at a clinical level which could inhibit adoption as a co-pilot or assistive tool to oncologists. Instead, CAD models should present clinically-relevant information based on the current way of working and allow the oncologist to asses each point to finally decide on patient treatment options. Despite the remarkable progress in utilizing deep learning models for PDAC segmentation, the achieved accuracies are still relatively low and may not be adequate for determining PDAC resectability.
%These features are not presented in an interpretable manner or aligned with clinical protocol for assessing tumor-vessel involvement. and their appearance on multi-phase contrast enhanced CT.
\section{Methods}
\subsection{Data Collection}\label{sec:data_collection}
This retrospective single-center research study investigate PDAC resectability in 99 patients specifically located in the pancreatic head. Determined by radiological assessment, a group of 50 patients have PDAC without vascular involvement, while 49 patients have PDAC with potential involvement of critical adjacent vasculature. We employ contrast-enhanced CT images obtained from the Catharina hospital. Each patient underwent a multi-phase pancreatic protocol CT scan, including (at least) the portal-venous phase, parenchymal phase, arterial phase, or late liver phase. Consequently, a total of 195 CT scans were included in our analysis. Prior to conducting the research, all CT scans were meticulously annotated. Under supervision of an expert abdominal radiologist, a surgical resident manually annotated all the relevant anatomical structures at voxel-level, including the tumor, pancreas, pancreatic duct~(PD), common bile duct~(CBD), aorta, superior mesenteric artery~(SMA), celiac trunk, hepatic artery, splenic artery, splenic vein, superior mesenteric vein~(SMV), portal vein, gastroduodenal artery~(GA) and inferior vena cava. For model training purposes, we aggregated the different arteries into a single arterial structure and, similarly, all the veins into one venous structure. Determining tumor resectability requires careful consideration of tumor presence, size and its relationship with surrounding anatomical structures. Particularly, the extent of contact between the tumor and neighboring veins and arteries plays a crucial role. This degree of contact is typically computed after the clinican made segmentation delineations where each of the structures are (or could be based on their best knowledge). Consequently, CT voxels have the potential to belong to multiple structures simultaneously. Figure~\ref{fig:involvement} illustrates an example along with corresponding ground-truth annotations of the involvement. Due to the inherent ambiguity in the data and low contrast in some phases, segmentations and the derived tumor-vessel involvement varies between subsequent scans of the same patient. As such, reported results are on a per scan basis. 
%The dataset comprises 90~scans without involvement, 9~cases with only arterial involvement, 61~with venous involvement and 35~with both arterial and venous involvement.
%Eligible participants were aged 18 years or older and had undergone surgical treatment for pancreatic head cancer at the Catharina Hospital Eindhoven. To ensure comprehensive data collection, both a surgical report and a complete pathology report were required for each patient.
%Catharina Hospital Eindhoven
%IntelliSpace Portal, a software package provided by Philips Healthcare in the Netherlands, was employed for the annotation process.
%Therefore, substantial effort was dedicated to annotating not only the tumor but also the surrounding features.
%  To ensure comprehensive information capture, we adopted a strategy of separately annotating and storing each structure, allowing for maximum data utilization.
\subsection{Segmentation models}
\begin{figure*}[t]
\begin{centering}
    \includegraphics[width=1.0\linewidth]{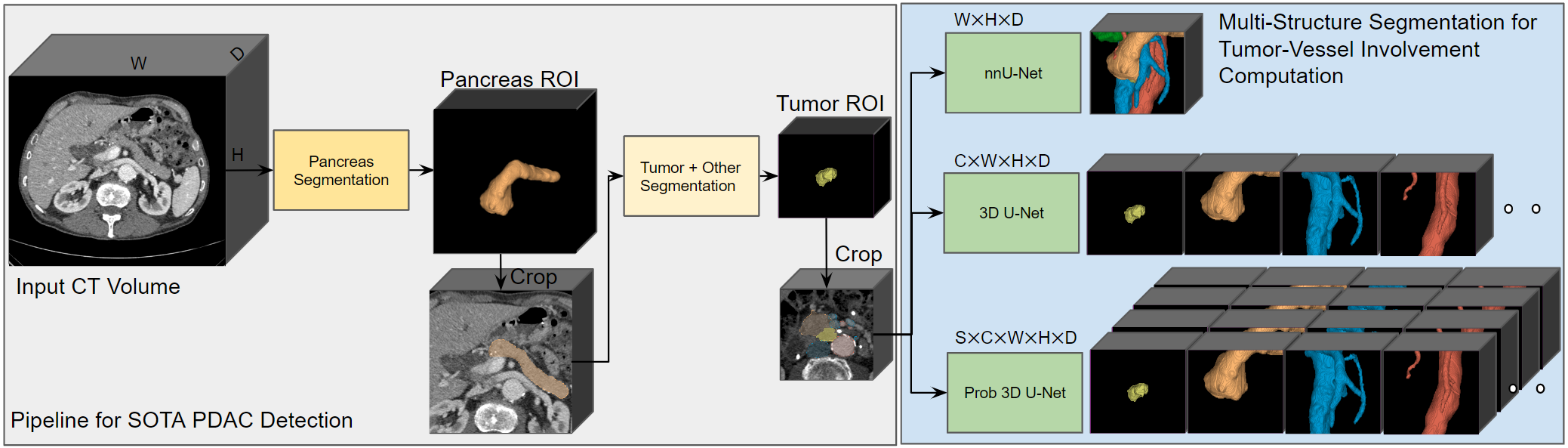}
\end{centering}
\vspace{-0.5cm}
  \caption{Workflow illustration for tumor segmentation from a CT scan. From the CT scan, the pancreas is automatically segmented and cropped for automated tumor segmentation~\cite{viviers_improved_2022,alves_fully_2022}. With the tumor detected and segmented, another crop is taken around the tumor and provided to our three models to determine tumor-vessel involvement.}
  \label{fig:models}
\end{figure*}

This research employs three segmentation models to segment the tumor and surrounding anatomy. We train the (1)~3D nnU-Net to automatically segment the structures of interest in 3D. The six different structures are layered from the least to most important: pancreas, common bile duct, pancreatic duct, arteries, veins, tumor. To determine overlap, a 7th and 8th pseudo structure is created for the tumor-artery and tumor-vein overlap. A custom 3D U-Net is developed (2)~to segment the structures in multi-channel 3D, alleviating the need for pre-computed pseudo labels and enabling direct overlap prediction. The model is set up to be identical to that of the default nnU-Net, except for a final sigmoid activation, instead of softmax probabilities in the nnU-Net, that enable mutually independent class predictions. This approach was also chosen to have a fair indication of the effect of our novel overlap loss~(\ref{eq:oll}). The (3)~Probabilistic 3D U-Net follows the same segmentation scheme as the 3D U-Net and is utilized to express the aleatoric uncertainty in the structures of interest by presenting multiple plausible segmentation hypotheses. In Figure~\ref{fig:models}, these three approaches are showcased along with the initial tumor detection pipeline. 

%

%\subsection{Experiments}

%To test our approach and the extent of the influence of the external secondary features in the tumor detection, the following experiments are conducted. (1)~We start by setting the baseline at detecting a tumor using only the CT scan. This baseline is set using the popular nnU-Net~\cite{nnUnet} (Full-Resolution 3D) and a custom 3D U-Net. (2)~In a follow-up experiment, we add the detailed segmentation maps of the pancreas and ducts to the CT scan, concatenated channel-wise. The same custom 3D U-Net is trained to segment the tumor, but now with this additional information derived from the radiologist. (3)~As an ablation experiment, we replace the segmentation maps of the ducts with a Boolean input. The pancreatic and common bile ductal 3D volumes are replaced with unity values if they are dilated or not. (4) Finally, we apply the models (using the CT scan and detailed segmentation maps), trained and validated on the three data folds of our dataset, to the Medical Decathlon Dataset as test set.

\subsection{Data Preparation \& Training Details}\label{Sec:DataPrep}

The data for resectability prediction is prepared according to the workflow depicted in Figure~\ref{fig:involvement}. In Section~\ref{sec:related_work}, various methods are presented that achieve high PDAC detection accuracy with reasonable segmentation accuracy. As such, we continue by cropping around the tumor center. This is implemented based on the ground-truth tumor labels, however, in practice this is performed by a prior segmentation model. We crop the CT scan and corresponding labels of the tumor, pancreas, pancreatic duct, common bile duct and an aggregate of all the arteries and veins (as mentioned in Section~\ref{sec:data_collection}). For the nnU-Net implementation, two additional overlapping pseudo labels are created. The dataset is resampled to the mean dataset size ([1mm, 0.67mm, 0.67mm] in the $z,y,x$-axes) and cropped to [64, 128, 128] voxels in the $z,y,x$-axes, respectively.

% $\mathbf{x} \in \mathcal{X}$ which resides in $\mathbb{R}^{C \times Z \times W \times H}$ is the 4D volume of input data and $\mathbf{Y}_n \in \mathbb{R}^{ Z  \times W  \times H}$ is the 3D nnU-Net segmentation map, $\mathbf{Y}_n \in \mathbb{R}^{ C \times Z  \times W  \times H}$ is the 3D U-Net segmentation map and $\mathbf{Y}_n \in \mathbb{R}^{S \times  C \times Z  \times W  \times H}$ is the Probabilistic 3D U-Net output.

We perform a 70\%/15\%/15\% patient dataset train/validation/test split. The test split is chosen by a surgical resident to be representative of the distribution of the tumor size, location and involvement present in our dataset. From the 85\% train/validation data we perform threefold bootstrapping using random patient splits and report results on the validation and test splits. The full resolution 3D nnU-Net is employed as reported publicly~\footnote{\href{https://github.com/MIC-DKFZ/nnUNet}{https://github.com/MIC-DKFZ/nnUNet}} without any modifications. The custom 3D U-Net is implemented in PyTorch and extends on the work by Wolny~\etal~\cite{eLife}~\footnote{\href{https://github.com/wolny/pytorch-3dunet}{https://github.com/wolny/pytorch-3dunet}}. The Probabilistic 3D U-Net is adapted from the implementation by Viviers~\etal~\cite{prob3dunet} and available online~\footnote{\href{https://github.com/cviviers/prob\_3D\_segmentation}{https://github.com/cviviers/prob\_3D\_segmentation}}. During training, we only employ the loss function introduced in Eq.~(\ref{eq:combo_loss}) and all other U-Net model parameters are chosen to be consistent with that of the nnU-Net where possible. The U-Nets are 5 layers deep with 32, 64, 128, 256, and 320 filters at each layer, respectively. A cosine annealing strategy is employed to modulate the Probabilistic 3D U-Net elbo beta between 1 and 10. All the models were trained for 1000 epochs with the Adam optimizer and a linear decaying learning rate scheduler. During training, the model weights with the best validation performance are chosen. %The implementations will be made publicly available for reproducibility~\footnote{\href{TBD}{TBD}}.

Recent advancements in semantic segmentation for medical applications have demonstrated the effectiveness of combining binary cross-entropy~(BCE) and Dice loss functions to enhance performance~\cite{nnUnet}. The cross-entropy loss is proficient in capturing global context and penalizing misclassifications, while the Dice loss emphasizes spatial overlap and similarity. Although optimizing these objectives contributes to determining the overlap between tumors and vessels, we propose a loss function, called the Overlap Loss~(OLL). Let $\mathbf{x}\in\mathcal{X}$ and $\mathbf{y}\in\mathcal{Y}$ be random variables taking values in $\mathbb{R}^{Z \times H \times W}$ and $\mathbb{R}^{C \times Z \times H \times W}$, and representing the input images ($Z$ image depth, $H$-height and $W$-width) and ground-truth masks, respectively. 
% We denote the dataset as the set of image-segmentation pairs $\mathcal{D} = \{(\mathbf{x}_1, \mathbf{y}_1), ...(\mathbf{x}_N, \mathbf{y}_N)\}$, with $N$ being the dataset size. 
We define random variables $\mathbf{T}\in\mathcal{T}$, $\mathbf{A}\in\mathcal{A}$ and $\mathbf{V}\in\mathcal{V}$, representing tumor, artery and vein segmentation masks, respectively, which are elements in subsets of $\mathcal{Y}$. 
% \begin{equation}
%     \mathcal{S}=\{\mathcal{T}, \mathcal{A}, \mathcal{V}\} \hspace{10pt} \wedge \hspace{10pt} s \subseteq \mathcal{Y} \; \forall \; s \in\mathcal{S}.
% \end{equation}
Let the pseudo overlap labels $\alpha$ and $\nu$ for the tumor-artery and tumor-vein pairs be defined as 
\begin{equation}
        \alpha = \mathbf{T} \odot \mathbf{A},\;  \nu = \mathbf{T} \odot \mathbf{V},
\end{equation}
where $\odot$ implies the element-wise product. Then, the OLL is denoted by
\begin{equation}\label{eq:oll}
     \begin{aligned}
        H_{o} = H(\hat{\alpha}, \alpha) + H(\hat{\nu}, \nu), 
    \end{aligned}
\end{equation}
where $H$ is the element-wise BCE. Hat notation is utilized to differentiate sigmoid-activated logit predictions from ground-truth masks. It is important to note that sigmoid activation precedes the creation of the pseudo labels.
%
% \begin{equation}\label{eq:oll}
%  \begin{aligned}
% \text{OLL} = \text{CE}\Bigl( \sigma(\text{T}_\text{L}) \odot \sigma(\text{A}_\text{L}) , (\text{T}_\text{GT}) \odot (\text{A}_\text{GT})\Bigr) \\
% + \text{CE} \Bigl( \sigma(\text{T}_\text{L}) \odot \sigma(\text{V}_\text{L}), (\text{T}_\text{GT}) \odot (\text{V}_\text{GT}) \Bigr) ,
% \end{aligned}
% \end{equation}
%

% The OLL calculates a pseudo overlap label for tumor~(T$_{\text{GT}}$)-artery~(A$_{\text{GT}}$) and tumor~(T$_{\text{GT}}$)-vein~(V$_{\text{GT}}$) pairs and compares them to the predicted overlapping structures (Artery logits~(A$_\text{L}$), Vein logits(V$_\text{L}$) and Tumor logits~(T$_\text{L}$), where $\sigma (\cdot) $ denotes the sigmoid function). 
The OLL directly aims to optimize the predictions of the overlapping structures. Although the objective to accurately predict the degrees of involvement is computed based on contact/adjacent pixels (see Section~\ref{sec:vessel_involvement}), we conjure that accurate overlap prediction will result in precise delineation of the contact area. By introducing this direct objective, we anticipate further improvements in the accuracy of segmentation results in the areas of interest and subsequent analysis. The complete training objective (CLL) for the 3D U-Net and the reconstruction loss for the Probabilistic 3D U-Net can then be formulated as
\begin{equation}\label{eq:combo_loss}
 \begin{aligned}
\text{CLL} = \alpha \times \Bigl[ \beta \times H(p,q) + (1 -\beta) \times \text{Dice}(p,q) \Bigr] \\
+ (1-\alpha) \times H_{o}(p,q), 
\end{aligned}
\end{equation}
where $\beta$ and $\alpha$ are weighting factors between the different loss components. Empirically, we found that $\beta$ = 0.5 and $\alpha$ = 0.8 works well.

% In all our experiments, the same crops, hyperparameters and augmentation techniques are used, with a hardware configuration based on a TITAN RTX GPU\footnote{Commercially available from Nvidia Corp., CA, USA}.
%cross-entropy loss, a batch size of 2, an Adam optimizer with an initial learning rate of $1\cdot10^{-4}$ and a weight decay of $1\cdot10^{-5}$. We use extensive data augmentation, consisting random rotation, elastic deformation, contrast adjustment, and additive Gaussian and Poisson Noise. BCE and Dice loss and $\alpha$ a weighting factor between the former mentioned loss and the OLL. 

\subsection{Computing \& Assessing Vessel Involvement}\label{sec:vessel_involvement}

The degree of tumor-vessel involvement is computed on a per 2D-axial slice basis. The contact area between the tumor and the vessel is calculated based on adjacent pixels, which is followed by calculating the vessel centroid and the length of each of the contact pixels' distance to the vessel center. The angle between each pixel and the center is calculated using the 4-quadrant arctangent function. The maximum and minimum angle are then used to determine the degree of involvement. The result can be observed in Figure~\ref{fig:involvement}. It is important to note that while clinicians do not have automated tools to perform this, the degree of involvement is assessed in a similar way. 

We introduce a classification metric of predicting involvement with the arteries and veins: if the resulting tumor and artery/vein segmentation predictions have involvement (degree $>$ 0), even at the wrong location compared to the GT, while the GT also has involvement somewhere, we consider it a true positive~(TP) prediction. If there is involvement prediction and the GT has no involvement, it is a false positive~(FP). In the case of no predicted involvement whatsoever and the GT also has no involvement, we consider it a true negative (TN) and vice-versa for false negatives~(FN). For the scan level sensitivity and specificity we follow the same procedure (if either the artery or vein involvement is a TP, then at scan level it is a TP and so forth). Additionally, we provide sensitivity and specificity result for the clinically relevant SMV, PV, SMA, Truncus. In this case, we remove the arterial and venous predictions with overlap with the pancreas and compare it to the GT aggregate of the SMV and PV (venous) or SMA and Truncus (arterial). The Dutch Pancreatic Cancer Group~(DPCG)~\cite{DPCG} classifies tumor resectablity based on the degrees of tumor-vessel contact. The tumor is resectable if SMV and PV has $\leq$ 90$^{\circ}$ contact, borderline resectable if the contact is between 90$^{\circ}$ and 270$^{\circ}$ and irresectable if the contact is $>$ 270$^{\circ}$. For the arterial vasculature it is borderline resectable for $\leq$ 90$^{\circ}$ contact and any amount of involvement more than 90$^{\circ}$ deems the tumor irresectable.

\subsection{The Effect of Ambiguity on Tumor-Vessel Involvement}\label{sec:uncertainty}
Accurate estimation of uncertainty is vital in image segmentation tasks to assess the reliability of the predicted segmentations. In this study, we compute the tumor-vessel involvement directly from the segmentations and, as such, any variation in the resulting segmentation can have a large impact on the involvement prediction and the extent (degree) thereof. We propose a comprehensive approach that ensembles different model folds and a probabilistic modeling approach to capture both epistemic and aleatoric uncertainty in the resulting segmentations.

To capture epistemic uncertainty, we construct an ensemble of our segmentation models. Each model in the ensemble is trained with different weight initializations and training dataset folds. By considering the disagreement among the ensemble members as samples from the model weight distribution, we can effectively capture the model's epistemic uncertainty regarding the true segmentation~\cite{gawlikowski2021survey,zhang2021dense}. The ensemble for 3D nnU-Nets and 3D U-Nets are thus capable of expressing the epistemic uncertainty. To address aleatoric uncertainty associated with ambiguity in the image data, we employ a probabilistic U-Net~~\cite{kohl2018probabilistic,valiuddin2021improving}. The probabilistic U-Net explicitly models the uncertainty within data by learning a lower-dimensional latent distribution of plausible variations in the output. This enables us to capture the inherent variability and ambiguity in voxel-level predictions~\cite{valiuddin2021improving}. By integrating the ensemble of probabilistic U-Nets, we obtain a holistic uncertainty estimation framework that captures both epistemic and aleatoric uncertainty. This combined uncertainty estimation approach enhances the interpretability and reliability of the segmentation results.

Practically, for the nnU-Net and 3D U-Net, we interpret the mean and standard deviation of the predicted segmentation probabilities across the 3 folds as the mean prediction and the epistemic uncertainty. The predicted probabilistic U-Net sigmoid probabilities can be writen as $\mathcal{Y} \in \mathbb{R}^{S \times C \times Z \times H \times W}$, where $S$ are the samples. Computing $\sigma(Y)$ is the aleatoric uncertainty and $\mu(\sigma(Y_0), \sigma(Y_1), \sigma(Y_2))$ is the mean aleatoric uncertainty across the three model folds. The epistemic uncertainty can be computed as $\sigma(\mu(Y_0), \mu(Y_1), \mu(Y_2))$. Figure~\ref{fig:seg_models_results} depicts the sum of the aleatoric and epistemic uncertainty for the probabilistic U-Net.

% The while an ensemble of the three probabilistic U-Net models indicates epistemic uncertainty. The standard deviation of all 16~samples across all three folds (48 predictions per image) captures both the epistemic and aleatoric uncertainty. The per voxel standard deviation of the 16~samples taken from the probabilistic U-Net express the aleatoric uncertainty.  The ensemble provides a robust estimation of uncertainty by leveraging the collective knowledge of multiple models. 

\section{Results \& Discussion}
\begin{figure}
\centering
    \includegraphics[width=0.6\linewidth]{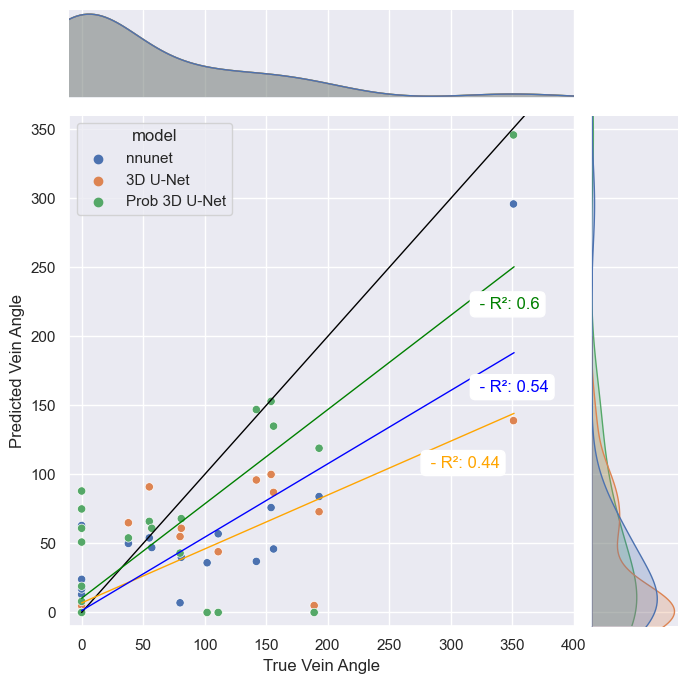}

  \caption{Maximum SMV or PV degrees of involvement.}
  \label{fig:vein_involvement_plot}
  \vspace{-0.5cm}
\end{figure}

\begin{figure*}
\centering
\resizebox{0.75\linewidth}{!}{
    \includegraphics[width=1.0\linewidth]{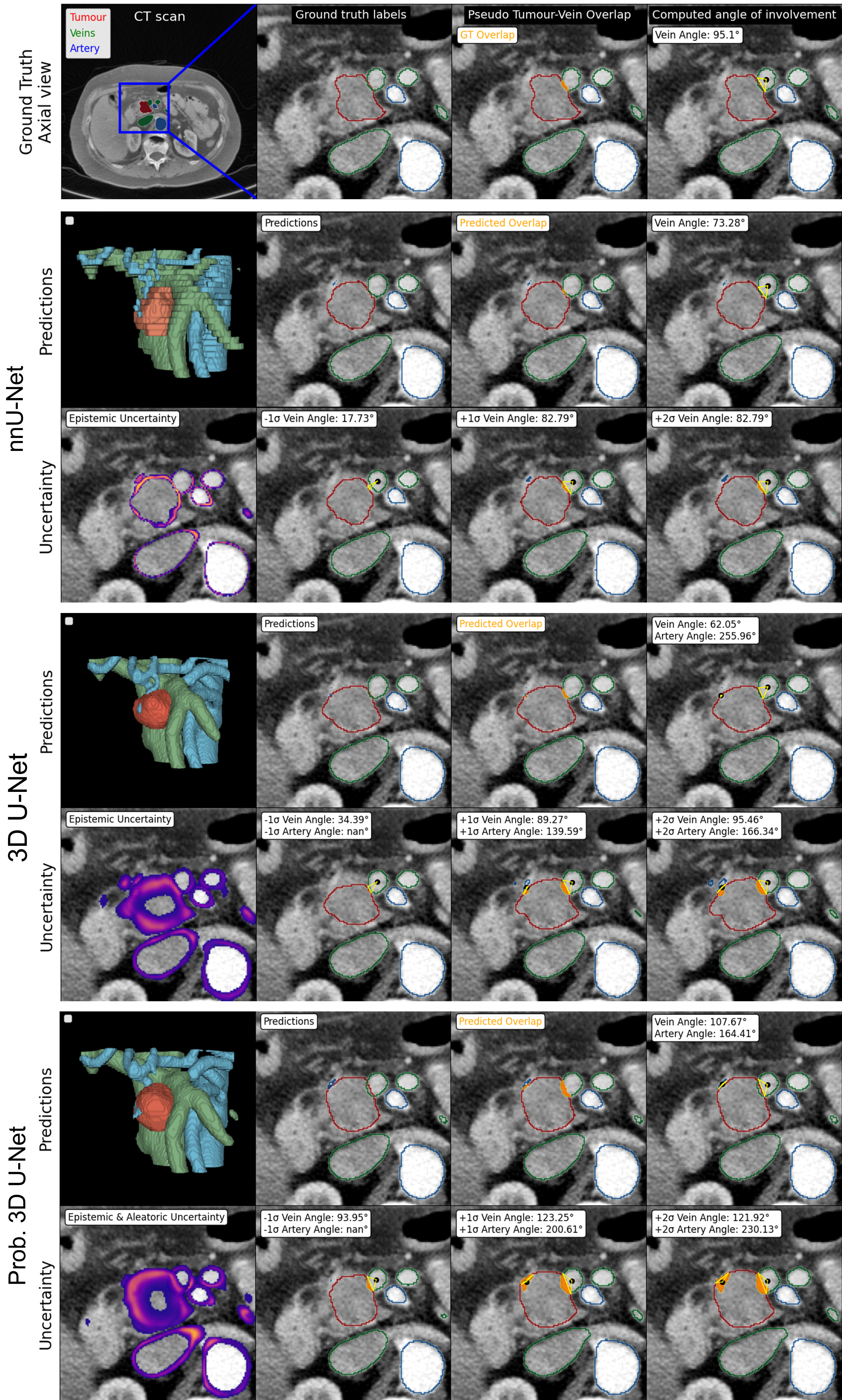}}

  \caption{Ground truth~(top) and predictions of the three models from a test set case.}
  \label{fig:seg_models_results}
\end{figure*}

% \subsection{Involvement Prediction}
The experimental results are listed in Table~\ref{tab:results},  Table~\ref{tab:r2_results} and Figure~\ref{fig:seg_models_results}. The mean and standard deviations are reported of the sensitivity, specificity and Dice (across all cases) on the validation sets, the three different models (from the folds) on the test set and an ensemble of the models' folds predictions on the test set. We do not include any results on segmentation performance of the pancreas, PD or CBD since it does not directly contribute to the tumor-vessel involvement focus of this study. As presented in Table~\ref{tab:r2_results} and Figure~\ref{fig:vein_involvement_plot}, we provide an $R^2$ score on how well the predicted maximum involvement correlates to the GT maximum involvement. 
\begin{table}
\centering
{\resizebox{1.0\linewidth}{!}{
\begin{tabular}{r|c|c|c} 
\toprule
 \textbf{Metric } \enspace &  \textbf{3D nnU-Net}  & \textbf{3D U-Net OLL} & \textbf{Prob. 3D U-Net OLL}  \\ 
\midrule
\multicolumn{3}{l}{Validation} \\
\midrule
Tumor Dice \enspace  &  \enspace$ 0.67 \pm 0.03$ \enspace & \enspace$ 0.65 \pm 0.02$ \enspace & \enspace$ 0.50 \pm 0.03$ \enspace  \\
Artery Dice  \enspace  &  \enspace$ 0.88 \pm 0.02$  \enspace & \enspace$ 0.83 \pm 0.03$ \enspace & \enspace$ 0.84 \pm 0.03$ \enspace  \\
Vein Dice \enspace  &\enspace $0.87 \pm 0.02$ \enspace & \enspace$ 0.85 \pm 0.03$ \enspace & \enspace$ 0.86\pm 0.02$ \enspace \\
Artery Overlap Dice \enspace  &\enspace $0.05 \pm 0.04$ \enspace& \enspace$ 0.04 \pm 0.03$ \enspace & \enspace$ 0.05\pm 0.03$ \enspace  \\
Vein Overlap Dice \enspace  &\enspace $0.16 \pm 0.08$ \enspace& \enspace$ 0.17 \pm 0.05$ \enspace & \enspace$ 0.14\pm 0.04$ \enspace  \\
Artery Sensitivity \enspace &\enspace $0.35 \pm 0.18$ \enspace& \enspace$ 0.48 \pm 0.17$ \enspace& \enspace$ 0.49\pm 0.12$ \enspace \\
Artery Specificity \enspace &\enspace $1.00 \pm 0.00$ \enspace& \enspace$ 0.90 \pm 0.08$ \enspace & \enspace$ 0.85\pm 0.05$ \enspace \\
Vein Sensitivity \enspace &\enspace $0.79 \pm 0.15$ \enspace& \enspace$ 0.86 \pm 0.13$ \enspace& \enspace$ 0.85\pm 0.05$ \enspace  \\
Vein Specificity \enspace &\enspace $0.87 \pm 0.11$ \enspace& \enspace$ 0.87 \pm 0.07$ \enspace& \enspace$ 0.73\pm 0.25$ \enspace \\
Scan Sensitivity \enspace &\enspace $0.81 \pm 0.14$ \enspace&\enspace$ 0.87\pm 0.11$ \enspace& \enspace$ 0.87\pm 0.03$ \enspace \\
Scan Specificity \enspace &\enspace $0.92 \pm 0.11$ \enspace& \enspace$ 0.85 \pm 0.03$ \enspace& \enspace$ 0.74\pm 0.2$ \enspace \\
\midrule
\multicolumn{3}{l}{Test} \\ 
\midrule
Tumor Dice \enspace  &  \enspace$ 0.65 \pm 0.01$ \enspace & \enspace$ 0.63 \pm 0.01$ \enspace & \enspace$ 0.49\pm 0.08$ \enspace   \\
Artery Dice  \enspace  &  \enspace$ 0.86 \pm 0.00$  \enspace & \enspace$ 0.86 \pm 0.01$ \enspace & \enspace$ 0.86\pm 0.01$ \enspace   \\
Vein Dice \enspace  &\enspace $0.90 \pm 0.00$ \enspace & \enspace$ 0.87 \pm 0.00$ \enspace & \enspace$ 0.88\pm 0.01$ \enspace  \\
Artery Overlap Dice \enspace  &\enspace $0.00 \pm 0.00$ \enspace& \enspace$ 0.02 \pm 0.01$ \enspace & \enspace$ 0.01\pm 0.00$ \enspace  \\
Vein Overlap Dice \enspace  &\enspace $0.08 \pm 0.01$ \enspace& \enspace$ 0.14 \pm 0.03$ \enspace & \enspace$ 0.12\pm 0.02$ \enspace \\
Artery Sensitivity \enspace &\enspace $0.23 \pm 0.05$ \enspace& \enspace$ 0.53 \pm 0.17$ \enspace& \enspace$ 0.05\pm 0.08$ \enspace \\
Artery Specificity \enspace &\enspace $0.94 \pm 0.02 $ \enspace &\enspace$ 0.91 \pm 0.04$ \enspace& \enspace$ 0.83\pm 0.08$ \enspace \\
Vein Sensitivity \enspace &\enspace $0.85 \pm 0.06$ \enspace& \enspace$ 0.81 \pm 0.09$ \enspace & \enspace$ 0.73\pm 0.08$ \enspace  \\
Vein Specificity \enspace &\enspace $0.77 \pm 0.06$ \enspace& \enspace$ 0.75 \pm 0.18$ \enspace& \enspace$ 0.60\pm 0.11$ \enspace \\
Scan Sensitivity \enspace &\enspace $0.81 \pm 0.00$ \enspace& \enspace$ 0.83 \pm 0.08$ \enspace & \enspace$ 0.77\pm 0.07$ \enspace \\
Scan Specificity \enspace &\enspace $0.74 \pm 0.07$ \enspace& \enspace$ 0.74 \pm 0.13$ \enspace& \enspace$ 0.67\pm 0.09$ \enspace \\
\midrule
\multicolumn{3}{l}{Test Ensemble} \\ 
\midrule
Tumor Dice \enspace  &  \enspace $\mathbf{0.66}$ \enspace & \enspace $\mathbf{0.66}$ \enspace & 0.56  \\
Artery Dice  \enspace  &  \enspace $\mathbf{0.86}$  \enspace & \enspace $\mathbf{0.86}$ \enspace & 0.87  \\
Vein Dice \enspace  &\enspace $\mathbf{0.91}$ \enspace & \enspace $0.88$ \enspace & 0.89 \\
Artery Overlap Dice \enspace  &\enspace $0.00$ \enspace& \enspace $\mathbf{0.01}$ \enspace & $\mathbf{0.01}$ \\
Vein Overlap Dice \enspace  &\enspace $0.07$ \enspace& \enspace $\mathbf{0.15}$ \enspace & 0.13 \\
Artery Sensitivity \enspace &\enspace $0.20$ \enspace& \enspace $0.30$ \enspace & $\mathbf{0.40}$\\
Artery Specificity \enspace &\enspace $0.91$ \enspace& \enspace $\mathbf{1.00}$ \enspace& 0.95\\
Vein Sensitivity \enspace &\enspace $0.81$ \enspace& \enspace $\mathbf{0.88}$ \enspace& 0.75\\
Vein Specificity \enspace &\enspace $0.81$ \enspace& \enspace $\mathbf{0.81}$ \enspace& 0.62\\
Scan Sensitivity \enspace &\enspace $0.81 $ \enspace& \enspace $\mathbf{0.88}$ \enspace& 0.79\\
Scan Specificity \enspace &\enspace $0.79$ \enspace& \enspace $\mathbf{0.86}$ \enspace& 0.72\\
\midrule
\multicolumn{3}{l}{Test Ensemble Predictions with only the SMV, PV, SMA, Truncus involvement} \\ 
\midrule
SMA or Truncus Sensitivity \enspace &\enspace $\mathbf{0.50}$ \enspace& \enspace $\mathbf{0.50}$ \enspace & $\mathbf{0.50}$\\
SMA or Truncus Specificity \enspace &\enspace $0.90$ \enspace& \enspace $\mathbf{0.93}$ \enspace& 0.90\\
SMV or PV Specificity \enspace &\enspace $\mathbf{0.92}$ \enspace& \enspace $\mathbf{0.92}$ \enspace& 0.77\\
SMV or PV Specificity \enspace &\enspace $0.79$ \enspace& \enspace $\mathbf{0.89}$ \enspace& 0.68\\
Scan Sensitivity. \enspace &\enspace $\mathbf{0.92} $ \enspace& \enspace $\mathbf{0.92}$ \enspace& 0.79\\
Scan Specificity \enspace &\enspace $0.79$ \enspace& \enspace $\mathbf{0.89}$ \enspace& 0.68\\
\bottomrule
\end{tabular}}
\caption{\textit{Segmentation and overlapping scores obtained with the 3D nnU-Net, 3D U-Net and Probabilistic 3D U-Net across 3 validation folds. These three models applied to the test set and an ensemble of these predictions.} \label{tab:results} }}
\end{table}

\begin{table}
\centering
{\resizebox{1.0\linewidth}{!}{
\begin{tabular}{r|c|c|c} 
\toprule

 \textbf{Metric } \enspace &  \textbf{3D nnU-Net}  & \textbf{3D U-Net} & \textbf{Prob. 3D U-Net}  \\ 
\midrule
\multicolumn{4}{l}{Validation} \\
\midrule
Artery $R^2$ \enspace &\enspace $-0.07 \pm 0.26$ \enspace& \enspace $-0.17 \pm 0.09$  \enspace & $-0.55 \pm 0.56$ \\
Vein $R^2$ \enspace &\enspace $0.34 \pm 0.39$ \enspace& \enspace $0.16 \pm 0.40$  \enspace & $-1.95 \pm 2.76$ \\
\midrule
\multicolumn{3}{l}{Test} \\ 
\midrule
Artery $R^2$ \enspace &\enspace $-0.24 \pm 0.01$ \enspace& \enspace $0.12 \pm 0.22$ \enspace & \enspace  $-0.24 \pm 0.17$  \enspace\\
Vein $R^2$ \enspace &\enspace $0.37 \pm 0.21$ \enspace& \enspace $0.42 \pm 0.13$ \enspace & \enspace $-0.04 \pm 0.44$ \enspace\\
\midrule
\multicolumn{4}{l}{Test Ensemble} \\ 
\midrule
Artery $R^2$ \enspace &\enspace $-0.27 $ \enspace& \enspace $-0.24 $ \enspace& -0.06\\
Vein $R^2$ \enspace &\enspace $0.52$ \enspace& 0.42 & 0.31\\
\midrule
\multicolumn{4}{l}{Test Ensemble with only the SMV, PV, SMA, Truncus involvement} \\ 
\midrule
Artery $R^2$ \enspace &  \enspace $-0.14 $ \enspace & \enspace $-0.01 $ \enspace& -7.53\\
Vein $R^2$ \enspace   &  \enspace $0.54$ \enspace & 0.44 & 0.60\\
\midrule
\multicolumn{4}{l}{Test Ensemble vascular criteria SMV, PV, SMA, Truncus involvement } \\ 
\midrule
0$^{\circ}$ $=$ Involvement \enspace                          & (19/19), (27/30)  & (19/19), (28/30) & (19/19), (27/30) \\
0$^{\circ}$ $<$ Involvement $\leq$ 90$^{\circ}$  \enspace  & (5/5), (1/1)  & (5/5), (1/1) & (5/5), (1/1)\\
90$^{\circ}$ $<$ Involvement $\leq$ 270$^{\circ}$ \enspace  & (0/7), (0/1)  & (2/7), (0/1) & (4/7), (0/1)\\
270$^{\circ}$ $<$ Involvement  \enspace                      &   (1/1), (0/0) & (0/1), (0/0) &  (1/1), (0/0)\\

\bottomrule
\end{tabular}}
\caption{Correlation between ground-truth and predicted involvement. Following the DPCG criteria~\cite{DPCG} the tumor-vein (first set of brackets) and tumor-artery (second set of brackets) involvement is categorized and assessed.\label{tab:r2_results} }}
\end{table}
We produce this plot by taking the maximum involvement angle at any slice (computed using the GT) and compare it with the maximum predicted angle of involvement. The maximum angle of involvement is one of the most important criteria used by clinicians in determining the treatment plan. Figure~\ref{fig:seg_models_results} showcases the performance of the three segmentation models on a scan from our test set. The particular model ensemble prediction, the overlapping structure (either predicted or derived) and computed degrees of involvement are presented. In the following row the associated uncertainty is showcased (computed as described in Section~\ref{sec:uncertainty}). The heat map is on a standard 0-0.5 scale and clipped below 0.01 to enable visualization of the background. The segmentations derived by subtracting 1, adding 1 and adding 2 voxel-level standard deviations are presented in the next columns to showcase the effect the uncertainty can have on the final tumor-vessel involvement prediction.

\emph{\textbf{Validation Results:}} In terms of segmentation accuracy, the nnU-Net outperforms the other models and achieves tumor, artery and vein Dice score of 0.67$\pm$0.03, 0.88$\pm$0.02 and 0.87$\pm$0.02. However, the overlap Dice scores for artery and vein are low, indicating difficulty in delineating these pseudo structures. The OLL used in the 3D U-Net hardly affected the Dice scores of the overlapping structures compared to the nnU-Net. Sensitivity and specificity values for artery and vein segmentation varies, with the vein sensitivity showing higher performance compared to artery sensitivity. However, the OLL significantly boosts both artery (0.48$\pm$0.17) and vein (0.86$\pm$0.07) sensitivity of the 3D U-Net at the cost of a few FP predictions.

\emph{\textbf{Test Results:}} The obtained segmentation accuracies align closely with the validation results. Specifically, the proposed OLL approach exhibits enhanced generalizability, with the 3D U-Net model showing slight improvements in the dice scores for overlapping structures. The artery overlap segmentation achieved a dice score of 0.02$\pm$0.01 compared to the baseline score of 0, while the vein segmentation yielded a Dice score of 0.14$\pm$0.03, surpassing the baseline score of 0.08$\pm$0.01. These low scores can be attributed to very small structures, large GT variability and general difficulty in accurate delineation due to lack of contrast. Sensitivity and specificity values for artery and vein segmentation on the test set are generally consistent with the validation results. However, it is worth noting that the 3D U-Net model displays larger standard deviations, indicating that some model folds agree well with the test set, while others demonstrate certain discrepancies.

\emph{\textbf{Test Ensemble:}} Predictions from the three folds are combined, resulting in a minor segmentation performance increase across all the models. The overlap Dice scores for artery and vein remain low, however, the 3D U-Net shows segmentations improvements over the nnU-Net that result in larger detection performance improvements for both involvement with the artery and vein.

\emph{\textbf{Degree of Involvement:}} In Table~\ref{tab:r2_results} and Figure~\ref{fig:vein_involvement_plot} it can be seen that there is moderate agreement between the predicted angle of involvement and the ground truth angle for the venous structures in the validation, test and test ensemble for the critical structures.  The final test ensemble showcases slightly better alignment for the Prob. U-Net~($R^2$ 0.60) for angles in the SMV or PV. Very little agreement for the degree of involvement with the arteries is shown. The test set only contains one case with involvement with the SMA and one with the Truncus. The remaining cases either have no involvement or involvement with the GA (as can be seen in Figure~\ref{fig:seg_models_results}) which can easily be ignored by removing arterial predictions with the pancreas. Table~\ref{tab:r2_results} showcases the degree of involvement according to the DPCG criteria. Despite not achieving perfect accuracy in predicting the exact angles (a challenging task to begin with), the models provide clinically-relevant evaluations that are accurate enough for practical use. All three models capture the extent of involvement almost perfectly for smaller degrees of involvement. It is worth mentioning that the models tend to underestimate the involvement for larger degrees of involvement, indicating a potential limitation in capturing extensive involvements accurately. This can be attributed to the absence of cases with extensive involvement in our dataset.
%, but also a potential useful attribute, since clinicians tend to overestimate the amount of involvement (Section~\ref{sec:intro})

\emph{\textbf{Test Ensemble Uncertainty:}} The models' uncertainties are presented in Figure~\ref{fig:seg_models_results} for a scan from our test set that, in this slice, appears to be borderline resectable (90$^{\circ}$ $<$Involvement$\leq$ 270$^{\circ}$) due to the involvement with the SMV. Incorporating the uncertainty in the segmentation predictions allows for a likelihood-based evaluation of the tumor-vessel degree of involvement. In the example, the nnU-Net underestimates the tumor size and involvement and is predicting a resectable tumor. With very defined uncertainty regions, taking uncertainty steps (-1 $\sigma$, +1 $\sigma$ and +2 $\sigma$) does not change the degree of involvement by much and therefore, the treatment strategy will not be affected. The 3D U-Net initially underestimates the involvement, but with +2 $\sigma$ steps, the borderline-resectable margin is crossed, indicating a potentially larger tumor with more involvement and the correct treatment approach. The Probabilistic 3D U-Net already predicts the correct response (borderline resectable) with a larger degree of involvement~(107.67$^{\circ}$). While the model is capable of expressing all the uncertainty, in this case it does not affect the predicted resectability.

Across the board we see good segmentation accuracies for the desired structures of veins, arteries and the pancreatic tumor. Although the deep learning models demonstrate promising segmentation results, there is still room for improvement, particularly in the tumor and capturing the area of overlap between the tumor and vessels which, we conjecture correlates with the contact area and ultimately the degrees of involvement. The lower tumor Dice can be connected to the lack of visual information, both in texture and contrast, in CT volumes concerning the tumor. This can be readily understood, since this overlap measurement is a secondary step after the primary step of obtaining sufficient segmentation accuracy of the individual components. Presenting the $R^2$ metric, the effect of OLL and valuable uncertainty estimates is a first attempt at accurately quantifying the amount of overlap and extent of tumor-vessel involvement. Obtaining a more accurate measurement and a metric that incorporates uncertainty in the involvement assessment is presently ongoing work. As for our primary objective, we obtain high classification accuracy from the segmentations, clearly predicting tumor-vessel (sensitivity 88\% and specificity 0.86\%) and tumor-critical vessel (sensitivity 92\% and specificity 0.89\%) involvement. These results are of high clinical value and very encouraging because it is achieved by mimicking the clinical way of working from deriving tumor-vessel contact based on the previously mentioned segmentation results.

\emph{\textbf{Limitations:}} These findings are based on a small dataset, particularly concerning tumor-artery involvement. Accurate assessment of tumor-vessel involvement heavily relies on precise segmentation, which needs to exactly match the annotations provided by experts. However, achieving such accuracy is challenging, considering the inherent ambiguity associated with tumor visibility on CT which is openly discussed among experienced clinicians. This work is one of the first to facilitate and contribute to this difficult problem that clinicians have to face on a daily basis with potentially severe patient consequences in decision-making. 

%In Table~\ref{tab:r2_results} and Figure~\ref{fig:vein_involvement_plot}, we observe a moderate level of agreement between the predicted degree of involvement and the ground truth degree for the venous structures. However, for the degree of involvement with the SMA or GA, little agreement is evident as can be noted from the low Artery $R^2$ score. Notably, there are significant variations among the different model folds for all three models. The final test ensemble demonstrates slightly improved alignment for the nnU-Net~($R^2$ 0.52). While the exact angles predicted by the models are not perfect, they are accurate enough to provide clinically-relevant evaluations. For larger degrees of involvement the models tend to underestimate the involvement.
%Table~\ref{tab:r2_results} presents the degree of involvement according to the DPCG criteria. 
%In addition, in practice the involvement with the GA is useful to know, but does not have high clinical impact as it is typically resected along with the tumor during surgery. 
\section{Conclusion}
This study is the first to present a workflow for acquiring a focused region of interest in pancreatic ductal adenocarcinoma (PDAC) CT scans, with the aim of predicting tumor-vessel involvement and tumor resectability. Three deep learning-based segmentation architectures (3D nnU-Net, a 3D U-Net with overlap loss~(OLL) and the Probabilistic 3D U-Net with OLL) have been implemented and evaluated for the automated segmentation of PDAC and important anatomical surrounding structures. These delineations are levered for the automated detection and computation of degree of involvement for tumor-vascular contact. In addition, we present the uncertainty captured by each of the models and show how it can affect the involvement prediction, providing clinicians with uncertainty intervals of involvement. The first stage of anatomical structure segmentation is achieved with remarkably high accuracies for veins (Dice 0.88), arteries (Dice 0.86), and, importantly, the pancreatic tumor (Dice 0.66) with the 3D U-Net, enabling the tumor-vessel involvement prediction. While precisely measuring the amount of involvement is still a challenge, utilizing the computed degrees of involvement with the critical structures to classify resectability shows compelling results. The presence of tumor involvement is determined with high sensitivity~(0.88) and specificity~(0.86), providing clinicians with a clear indication of involvement, paving the way for more informed decision-making capabilities for surgical interventions and personalized treatment strategies. 
\newpage

{\small
\bibliographystyle{ieee_fullname}
\bibliography{egbib}

\begin{thebibliography}{10}\itemsep=-1pt

\bibitem{alemi2016classification}
Farzad Alemi, Flavio~G Rocha, William~S Helton, Thomas Biehl, and Adnan
  Alseidi.
\newblock Classification and techniques of en bloc venous reconstruction for
  pancreaticoduodenectomy.
\newblock {\em HPB}, 18(10):827--834, 2016.

\bibitem{alves_fully_2022}
Natália Alves, Megan Schuurmans, Geke Litjens, Joeran~S. Bosma, John Hermans,
  and Henkjan Huisman.
\newblock Fully {Automatic} {Deep} {Learning} {Framework} for {Pancreatic}
  {Ductal} {Adenocarcinoma} {Detection} on {Computed} {Tomography}.
\newblock {\em Cancers}, 14(2), 2022.

\bibitem{asbun2020miami}
Horacio~J Asbun, Alma~L Moekotte, Frederique~L Vissers, Filipe Kunzler,
  Federica Cipriani, Adnan Alseidi, Michael~I D’Angelica, Alberto Balduzzi,
  Claudio Bassi, Bergthor Bj{\"o}rnsson, et~al.
\newblock The miami international evidence-based guidelines on minimally
  invasive pancreas resection.
\newblock {\em Annals of surgery}, 271(1):1--14, 2020.

\bibitem{chen_pancreatic_2023}
Po-Ting Chen, Tinghui Wu, Pochuan Wang, Dawei Chang, Kao-Lang Liu, Ming-Shiang
  Wu, Holger~R. Roth, Po-Chang Lee, Wei-Chih Liao, and Weichung Wang.
\newblock Pancreatic cancer detection on ct scans with deep learning: A
  nationwide population-based study.
\newblock {\em Radiology}, 306(1):172--182, 2023.
\newblock PMID: 36098642.

\bibitem{3dunet}
{\"O}zg{\"u}n {\c{C}}i{\c{c}}ek, Ahmed Abdulkadir, Soeren~S. Lienkamp, Thomas
  Brox, and Olaf Ronneberger.
\newblock 3d u-net: Learning dense volumetric segmentation from sparse
  annotation.
\newblock In Sebastien Ourselin, Leo Joskowicz, Mert~R. Sabuncu, Gozde Unal,
  and William Wells, editors, {\em Medical Image Computing and
  Computer-Assisted Intervention -- MICCAI 2016}, pages 424--432, Cham, 2016.
  Springer International Publishing.

\bibitem{De_la_Santa2014-jz}
Luis~Gij{\'o}n de~la Santa, Jos{\'e} Antonio~P{\'e}rez Retortillo,
  Ainhoa~Camarero Miguel, and Lea~Marie Klein.
\newblock Radiology of pancreatic neoplasms: An update.
\newblock {\em World J. Gastrointest. Oncol.}, 6(9):330--343, Sept. 2014.

\bibitem{de2014radiology}
Luis~Gij{\'o}n de~la Santa, Jos{\'e} Antonio~P{\'e}rez Retortillo,
  Ainhoa~Camarero Miguel, and Lea~Marie Klein.
\newblock Radiology of pancreatic neoplasms: An update.
\newblock {\em World journal of gastrointestinal oncology}, 6(9):330, 2014.

\bibitem{gawlikowski2021survey}
Jakob Gawlikowski, Cedrique Rovile~Njieutcheu Tassi, Mohsin Ali, Jongseok Lee,
  Matthias Humt, Jianxiang Feng, Anna Kruspe, Rudolph Triebel, Peter Jung,
  Ribana Roscher, et~al.
\newblock A survey of uncertainty in deep neural networks.
\newblock {\em arXiv preprint arXiv:2107.03342}, 2021.

\bibitem{Hidalgo2010-sp}
Manuel Hidalgo.
\newblock Pancreatic cancer.
\newblock {\em N. Engl. J. Med.}, 362(17):1605--1617, Apr. 2010.

\bibitem{nnUnet}
Fabian Isensee, Paul~F Jaeger, Simon A~A Kohl, Jens Petersen, and Klaus~H
  Maier-Hein.
\newblock {nnU-Net}: a self-configuring method for deep learning-based
  biomedical image segmentation.
\newblock {\em Nature Methods}, 18(2):203--211, Feb. 2021.

\bibitem{joo2019preoperative}
Ijin Joo, Jeong~Min Lee, Eun~Sun Lee, Jee-Young Son, Dong~Ho Lee, Su~Joa Ahn,
  Won Chang, Sang~Min Lee, Hyo-Jin Kang, and Hyun~Kyung Yang.
\newblock Preoperative ct classification of the resectability of pancreatic
  cancer: interobserver agreement.
\newblock {\em Radiology}, 293(2):343--349, 2019.

\bibitem{kohl2018probabilistic}
Simon~AA Kohl, Bernardino Romera-Paredes, Clemens Meyer, Jeffrey De~Fauw,
  Joseph~R Ledsam, Klaus~H Maier-Hein, SM Eslami, Danilo~Jimenez Rezende, and
  Olaf Ronneberger.
\newblock A probabilistic u-net for segmentation of ambiguous images.
\newblock {\em arXiv preprint arXiv:1806.05034}, 2018.

\bibitem{lermite2013complications}
Emilie Lermite, Daniele Sommacale, Tullio Piardi, Jean-Pierre Arnaud, Alain
  Sauvanet, Cornelis~HC Dejong, and Patrick Pessaux.
\newblock Complications after pancreatic resection: diagnosis, prevention and
  management.
\newblock {\em Clinics and research in hepatology and gastroenterology},
  37(3):230--239, 2013.

\bibitem{Liu_deep_2020}
Kao-Lang Liu, Tinghui Wu, Po-Ting Chen, Yuhsiang~M Tsai, Holger Roth,
  Ming-Shiang Wu, Wei-Chih Liao, and Weichung Wang.
\newblock Deep learning to distinguish pancreatic cancer tissue from
  non-cancerous pancreatic tissue: a retrospective study with cross-racial
  external validation.
\newblock {\em The Lancet Digital Health}, 2(6):e303--e313, June 2020.

\bibitem{liu_establishment_2019}
Shang-Long Liu, Shuo Li, Yu-Ting Guo, Yun-Peng Zhou, Zheng-Dong Zhang, Shuai
  Li, and Yun Lu.
\newblock Establishment and application of an artificial intelligence diagnosis
  system for pancreatic cancer with a faster region-based convolutional neural
  network.
\newblock {\em Chinese Medical Journal}, 132(23), 2019.

\bibitem{ma_construction_2020}
Han Ma, Zhong-Xin Liu, Jing-Jing Zhang, Feng-Tian Wu, Cheng-Fu Xu, Zhe Shen,
  Chao-Hui Yu, and You-Ming Li.
\newblock Construction of a convolutional neural network classifier developed
  by computed tomography images for pancreatic cancer diagnosis.
\newblock {\em World Journal of Gastroenterology}, 26(34):5156--5168, Sept.
  2020.

\bibitem{mahmoudi2022segmentation}
Tahereh Mahmoudi, Zahra~Mousavi Kouzahkanan, Amir~Reza Radmard, Raheleh Kafieh,
  Aneseh Salehnia, Amir~H Davarpanah, Hossein Arabalibeik, and Alireza
  Ahmadian.
\newblock Segmentation of pancreatic ductal adenocarcinoma (pdac) and
  surrounding vessels in ct images using deep convolutional neural networks and
  texture descriptors.
\newblock {\em Scientific Reports}, 12(1):3092, 2022.

\bibitem{patra_abstract_2021}
Anurima Patra, Korfiatis Panagiotis, Garima Suman, Ananya Panda, Sushil~Kumar
  Garg, and Ajit Goenka.
\newblock Abstract {PO}-084: {Automated} detection of pancreatic ductal
  adenocarcinoma ({PDAC}) on {CT} scans using artificial intelligence ({AI}):
  {Impact} of inclusion of automated pancreas segmentation on the accuracy of
  {3D}-convolutional neural network ({CNN}).
\newblock {\em Clinical Cancer Research}, 27(5\_Supplement):PO--084, Mar. 2021.
\newblock Num Pages: PO-084.

\bibitem{Rahib2014-dh}
Lola Rahib, Benjamin~D Smith, Rhonda Aizenberg, Allison~B Rosenzweig, Julie~M
  Fleshman, and Lynn~M Matrisian.
\newblock Projecting cancer incidence and deaths to 2030: the unexpected burden
  of thyroid, liver, and pancreas cancers in the united states.
\newblock {\em Cancer Res.}, 74(11):2913--2921, June 2014.

\bibitem{si_fully_2021}
Ke Si, Ying Xue, Xiazhen Yu, Xinpei Zhu, Qinghai Li, Wei Gong, Tingbo Liang,
  and Shumin Duan.
\newblock Fully end-to-end deep-learning-based diagnosis of pancreatic tumors.
\newblock {\em Theranostics}, 11(4):1982--1990, 2021.
\newblock Publisher: Ivyspring International Publisher.

\bibitem{tempero2021pancreatic}
Margaret~A Tempero, Mokenge~P Malafa, Mahmoud Al-Hawary, Stephen~W Behrman,
  Al~B Benson, Dana~B Cardin, E~Gabriela Chiorean, Vincent Chung, Brian Czito,
  Marco Del~Chiaro, et~al.
\newblock Pancreatic adenocarcinoma, version 2.2021, nccn clinical practice
  guidelines in oncology.
\newblock {\em Journal of the National Comprehensive Cancer Network},
  19(4):439--457, 2021.

\bibitem{tran2014radiographic}
Hop~S Tran~Cao, Alpana Balachandran, Huamin Wang, Graciela~M
  Nogueras-Gonz{\'a}lez, Christina~E Bailey, Jeffrey~E Lee, Peter~WT Pisters,
  Douglas~B Evans, Gauri Varadhachary, Christopher~H Crane, et~al.
\newblock Radiographic tumor-vein interface as a predictor of intraoperative,
  pathologic, and oncologic outcomes in resectable and borderline resectable
  pancreatic cancer.
\newblock {\em Journal of Gastrointestinal Surgery}, 18:269--278, 2014.

\bibitem{Treadwell2016-we}
Jonathan~R Treadwell, Hanna~M Zafar, Matthew~D Mitchell, Kelley Tipton, Ursina
  Teitelbaum, and Jane Jue.
\newblock Imaging tests for the diagnosis and staging of pancreatic
  adenocarcinoma: A meta-analysis.
\newblock {\em Pancreas}, 45(6):789--795, July 2016.

\bibitem{tummala2011imaging}
Pavan Tummala, Omer Junaidi, and Banke Agarwal.
\newblock Imaging of pancreatic cancer: An overview.
\newblock {\em Journal of gastrointestinal oncology}, 2(3):168, 2011.

\bibitem{Vaiyapuri_intelligent_2022}
Thavavel Vaiyapuri, Ashit~K. Dutta, I.~S.~Hephzi Punithavathi, P. Duraipandy,
  Saud~S. Alotaibi, Hadeel Alsolai, Abdullah Mohamed, and Hany Mahgoub.
\newblock Intelligent {Deep}-{Learning}-{Enabled} {Decision}-{Making} {Medical}
  {System} for {Pancreatic} {Tumor} {Classification} on {CT} {Images}.
\newblock {\em Healthcare}, 10(4), 2022.

\bibitem{valiuddin2021improving}
MM Valiuddin, Christiaan~GA Viviers, Ruud~JG van Sloun, and Fons van~der
  Sommen.
\newblock Improving aleatoric uncertainty quantification in multi-annotated
  medical image segmentation with normalizing flows.
\newblock pages 75--88. Springer, 2021.

\bibitem{versteijne2017considerable}
Eva Versteijne, Oliver~J Gurney-Champion, Astrid van~der Horst, Eelco Lens,
  M~Willemijn Kolff, Jeroen Buijsen, Gati Ebrahimi, Karen~J Neelis, Coen Rasch,
  Jaap Stoker, et~al.
\newblock Considerable interobserver variation in delineation of pancreatic
  cancer on 3dct and 4dct: a multi-institutional study.
\newblock {\em Radiation Oncology}, 12(1):1--10, 2017.

\bibitem{Preopanc}
Eva Versteijne, Eelco Lens, Astrid van~der Horst, Arjan Bel, Jorrit Visser,
  Cornelis J~A Punt, Mustafa Suker, Casper H~J van Eijck, and Geertjan van
  Tienhoven.
\newblock Quality assurance of the {PREOPANC} trial (2012-003181-40) for
  preoperative radiochemotherapy in pancreatic cancer : The dummy run.
\newblock {\em Strahlenther. Onkol.}, 193(8):630--638, Aug. 2017.

\bibitem{DPCG}
Eva Versteijne, Casper H~J van Eijck, Cornelis J~A Punt, Mustafa Suker,
  Aeilko~H Zwinderman, Miriam A~C Dohmen, Karin B~C Groothuis, Oliver R~C
  Busch, Marc G~H Besselink, Ignace H J~T de Hingh, Albert~J Ten~Tije, Gijs~A
  Patijn, Bert~A Bonsing, Judith de Vos-Geelen, Joost~M Klaase, Sebastiaan
  Festen, Djamila Boerma, Joris~I Erdmann, I~Quintus Molenaar, Erwin van~der
  Harst, Marion~B van~der Kolk, Coen R~N Rasch, Geertjan van Tienhoven, and
  {Dutch Pancreatic Cancer Group (DPCG)}.
\newblock Preoperative radiochemotherapy versus immediate surgery for
  resectable and borderline resectable pancreatic cancer ({PREOPANC} trial):
  study protocol for a multicentre randomized controlled trial.
\newblock {\em Trials}, 17(1):127, Mar. 2016.

\bibitem{viviers_improved_2022}
Christiaan G.~A. Viviers, Mark Ramaekers, Peter H.~N. de With, Dimitrios
  Mavroeidis, Joost Nederend, Misha Luyer, and Fons van~der Sommen.
\newblock Improved {Pancreatic} {Tumor} {Detection} by {Utilizing}
  {Clinically}-{Relevant} {Secondary} {Features}.
\newblock In Sharib Ali, Fons van~der Sommen,
  Bart{\textbackslash}lomiej~W{\textbackslash}ladys{\textbackslash}law
  Papie{\textbackslash}.z, Maureen van Eijnatten, Yueming Jin, and Iris
  Kolenbrander, editors, {\em Cancer {Prevention} {Through} {Early}
  {Detection}}, pages 139--148, Cham, 2022. Springer Nature Switzerland.

\bibitem{prob3dunet}
Christiaan G.~A. Viviers, M.~M.~Amaan Valiuddin, Peter H.~N. de With, and Fons
  van~der Sommen.
\newblock {Probabilistic 3D segmentation for aleatoric uncertainty
  quantification in full 3D medical data}.
\newblock In Khan~M. Iftekharuddin and Weijie Chen, editors, {\em Medical
  Imaging 2023: Computer-Aided Diagnosis}, volume 12465, page 124651I.
  International Society for Optics and Photonics, SPIE, 2023.

\bibitem{wang_maff_2023}
Heng Wang, Zhongyi Wu, Fei Wang, Wenting Wei, Kezhen Wei, and Zhaobang Liu.
\newblock {MAFF}: {Multi}-{Scale} and {Self}-{Adaptive} {Attention} {Feature}
  {Fusion} {Network} for {Pancreatic} {Lesion} {Detection} in {PET} / {CT}
  {Images}.
\newblock In {\em Proceedings of the 2022 6th {International} {Conference} on
  {Electronic} {Information} {Technology} and {Computer} {Engineering}},
  {EITCE} '22, pages 1412--1419, New York, NY, USA, 2023. Association for
  Computing Machinery.
\newblock event-place: Xiamen, China.

\bibitem{wang_learning_2021}
Yan Wang, Peng Tang, Yuyin Zhou, Wei Shen, Elliot~K. Fishman, and Alan~L.
  Yuille.
\newblock Learning {Inductive} {Attention} {Guidance} for {Partially}
  {Supervised} {Pancreatic} {Ductal} {Adenocarcinoma} {Prediction}.
\newblock {\em IEEE Transactions on Medical Imaging}, 40(10):2723--2735, 2021.

\bibitem{wei_multidomain_2023}
Wenting Wei, Guorong Jia, Zhongyi Wu, Tao Wang, Heng Wang, Kezhen Wei, Chao
  Cheng, Zhaobang Liu, and Changjing Zuo.
\newblock A multidomain fusion model of radiomics and deep learning to
  discriminate between {PDAC} and {AIP} based on {18F}-{FDG} {PET}/{CT} images.
\newblock {\em Japanese Journal of Radiology}, 41(4):417--427, Apr. 2023.

\bibitem{eLife}
Adrian Wolny, Lorenzo Cerrone, Athul Vijayan, Rachele Tofanelli, Amaya~Vilches
  Barro, Marion Louveaux, Christian Wenzl, Sören Strauss, David
  Wilson-Sánchez, Rena Lymbouridou, Susanne~S Steigleder, Constantin Pape,
  Alberto Bailoni, Salva Duran-Nebreda, George~W Bassel, Jan~U Lohmann, Miltos
  Tsiantis, Fred~A Hamprecht, Kay Schneitz, Alexis Maizel, and Anna Kreshuk.
\newblock Accurate and versatile 3d segmentation of plant tissues at cellular
  resolution.
\newblock {\em eLife}, 9:e57613, jul 2020.

\bibitem{Yao2023-mp}
Jiawen Yao, Kai Cao, Yang Hou, Jian Zhou, Yingda Xia, Isabella Nogues, Qike
  Song, Hui Jiang, Xianghua Ye, Jianping Lu, Gang Jin, Hong Lu, Chuanmiao Xie,
  Rong Zhang, Jing Xiao, Zaiyi Liu, Feng Gao, Yafei Qi, Xuezhou Li, Yang Zheng,
  Le Lu, Yu Shi, and Ling Zhang.
\newblock Deep learning for fully automated prediction of overall survival in
  patients undergoing resection for pancreatic cancer: A retrospective
  multicenter study.
\newblock {\em Ann. Surg.}, 278(1):e68--e79, July 2023.

\bibitem{zhang2021dense}
Jing Zhang, Yuchao Dai, Mochu Xiang, Deng-Ping Fan, Peyman Moghadam, Mingyi He,
  Christian Walder, Kaihao Zhang, Mehrtash Harandi, and Nick Barnes.
\newblock Dense uncertainty estimation.
\newblock {\em arXiv preprint arXiv:2110.06427}, 2021.

\bibitem{zhang_novel_2020}
Zhengdong Zhang, Shuai Li, Ziyang Wang, and Yun Lu.
\newblock A {Novel} and {Efficient} {Tumor} {Detection} {Framework} for
  {Pancreatic} {Cancer} via {CT} {Images}.
\newblock In {\em 2020 42nd {Annual} {International} {Conference} of the {IEEE}
  {Engineering} in {Medicine} \& {Biology} {Society} ({EMBC})}, pages
  1160--1164, 2020.

\bibitem{zhu_segmentation_2021}
Zhuotun Zhu, Yongyi Lu, Wei Shen, Elliot~K. Fishman, and Alan~L. Yuille.
\newblock Segmentation for {Classification} of {Screening} {Pancreatic}
  {Neuroendocrine} {Tumors}.
\newblock In {\em Proceedings of the {IEEE}/{CVF} {International} {Conference}
  on {Computer} {Vision} ({ICCV}) {Workshops}}, pages 3402--3408, Oct. 2021.

\bibitem{zhu_multi_scale_2019}
Zhuotun Zhu, Yingda Xia, Lingxi Xie, Elliot~K. Fishman, and Alan~L. Yuille.
\newblock Multi-scale {Coarse}-to-{Fine} {Segmentation} for {Screening}
  {Pancreatic} {Ductal} {Adenocarcinoma}.
\newblock In Dinggang Shen, Tianming Liu, Terry~M. Peters, Lawrence~H. Staib,
  Caroline Essert, Sean Zhou, Pew-Thian Yap, and Ali Khan, editors, {\em
  Medical {Image} {Computing} and {Computer} {Assisted} {Intervention} –
  {MICCAI} 2019}, pages 3--12, Cham, 2019. Springer International Publishing.

\end{thebibliography}
}

\end{document}